\documentclass[conference,letter,10pt]{IEEEtran}

\usepackage{amsmath,epsfig,cite,amsfonts,amssymb,psfrag,subfig}
\usepackage{multicol}

\def\no{\nonumber}

\newtheorem{theorem}{Theorem}
\newtheorem{pro}{Proposition}

\newtheorem{defi}{Definition}

\begin{document}
\title{Secret Key Agreement Using Correlated Sources over the Generalized Multiple Access Channel}

\author{\IEEEauthorblockN{Somayeh Salimi,   Mikael Skoglund}
\IEEEauthorblockA{School of Electrical Engineering and ACCESS Linnaeus Center\\
KTH Royal Institute of Technology, Stockholm, Sweden\\
\{somayen,skoglund\}@ee.kth.se}
 }
\maketitle

\vspace{-2.5cm}
\begin{abstract}
A secret key agreement setup between three users is considered in which each of the users 1 and 2 intends to share a secret key with user 3 and users 1 and 2 are eavesdroppers with respect to each other. The three users observe i.i.d. outputs of correlated sources and there is a generalized discrete memoryless multiple access channel (GDMMAC) from users 1 and 2 to user 3 for communication between the users. The secret key agreement is established using the correlated sources and the GDMMAC. In this setup, inner and outer bounds of the secret key capacity region are investigated. Moreover, for a special case where the channel inputs and outputs and the sources form Markov chains in some order, the secret key capacity region is derived. Also a Gaussian case is considered in this setup.
\end{abstract}


%


\section{Introduction}
\noindent Secret key sharing between two legitimate users in the presence of an eavesdropper was considered in \cite{Ahlswede} and \cite{Maurer} in the source and the channel model. In the source model, all the three users could observe i.i.d. outputs of correlated sources and there was a noiseless public channel with unlimited capacity for communication between the two legitimate users, through which all communications could be intercepted by the eavesdropper. In \cite{Ahlswede}, secret key capacity is characterized in the source model when the noiseless public channel is one-way. In \cite{helper}, secret key sharing in a source model with public channel of limited capacity is investigated in which there is a helper with access to source observations correlated with the others observations. In \cite{joint-source-channel}, the same problem of sharing a secret key is investigated in the source model where instead of the public channel, there is a noisy broadcast channel from the transmitter to the receiver and the eavesdropper. In \cite{salimi-new-source-model}, secret key sharing was considered in a framework where two users intended to share secret keys with a base station and the two users were eavesdroppers of each other's key. In \cite{salimi-new-source-model}, all the three users can observe i.i.d. outputs of correlated sources and there is a noiseless public channel with unlimited capacity from users 1 and 2 to user 3.

 Motivated by the above works, we consider secret key sharing in a framework similar to \cite{salimi-new-source-model} but in more conformity with real communication scenarios since the realization of public channel with unlimited capacity is not compatible with the noisy nature of wireless networks. Hence, instead of public channels with unlimited capacity, we consider noisy channels for communication between the users. Each of the users 1 and 2, as network users, intends to share a secret key with user 3, as a base station, and users 1 and 2 are the eavesdroppers with respect to each other. The three users have access to correlated sources and there is a generalized discrete memoryless multiple access channel (GDMMAC) to transmit the required information from users 1 and 2 to user 3. Users 1 and 2 govern the channel inputs of the GDMMAC and each of the three users receives his corresponding output from the channel. Each of the users 1 and 2 generates a secret key from his observations and sends the required information via the GDMMAC to user 3. Users 1 and 2 use the channel outputs to eavesdrop each other's key. For this setup, we derive an inner bound of the secret key capacity region in which a combination of wiretap codebook and secret key generation codebook along with Wyner-Ziv codebook is used. Furthermore, an explicit outer bound of the secret key capacity region is given and for a special case, the secret key capacity region is derived. Also, the problem is discussed in a Gaussian case.

The rest of the paper is organized as follows: in Section II, the proposed model is described. In Section III, our main results are given. A special case is investigated in Section IV. A Gaussian example is discussed in Section V. Conclusion and suggestions for future works are given in Section VI. Proofs of the theorems are presented in Appendices. Throughout the paper, a random variable is denoted with an upper case letter (e.g $X$) and its realization is denoted with the corresponding lower case letter (e.g., $x$). We use $X_{i}^{N}$ to indicate vector $(X_{i,1},X_{i,2},...,X_{i,N})$, and $X_{i,j}^{k}$ to indicate vector $(X_{i,j},X_{i,j+1},...,X_{i,k})$, where $i$ denotes the index of the corresponding user.

%
\section{Preliminaries}

\noindent Users 1, 2 and 3 observe correlated discrete memoryless sources $S_{1},S_{2}$ and $S_{3}$, respectively, with joint distribution $P_{S_{1},S_{2},S_{3}}$ in an i.i.d. manner. Furthermore, there is a GDMMAC with probability distribution $P_{Y_{1},Y_{2},Y_{3}|X_{1} ,X_{2}}$, independent of the sources, where users 1 and 2 govern the inputs $X_{1}$ and $X_{2}$, and then, outputs $Y_{1},Y_{2}$ and $Y_{3}$ are seen by users 1, 2 and 3, respectively. Each of users 1 and 2 intends to share a secret keys with user 3 where user 1 is the eavesdropper of user 2's secret key and vice versa.

For the secret key agreement, users 1 and 2 generate secret keys $K_{1}\!$ and $K_{2}\!$ as stochastic functions of the information available at them, i.e., $S_{1}^{N}$ and $S_{2}^{N}$, respectively. For $i\!=\!1,2,...,N$, the $i$-th channel inputs $X_{1,i}$ and $X_{2,i}$ are determined as stochastic mappings \small{$X_{1,i}\!\!=\!f_{1,i}(\!S_{1}^{N}\!)$ and $X_{2,i}\!\!=\!f_{2,i}(\!S_{2}^{N}\!)$ \normalsize{ by users 1 and 2, respectively. These inputs are sent over the GDMMAC and then, channel outputs $Y_{1,i},Y_{2,i}$ and $Y_{3,i}$ are observed by users 1, 2 and 3, respectively. User 3 computes estimates of keys $K_{1}\!$ and $K_{2}\!$ as:}
\begin{equation}
(\hat{K}_{1},\hat{K}_{2})=g(Y_{3}^{N},S_{3}^{N})
\end{equation}

\vspace{-.2cm}
\noindent where $g$ is a deterministic function. Users 1 and 2 use the channel outputs $Y_{1}^{N}$ and $Y_{2}^{N}$ to eavesdrop on each other's key, respectively. For simplicity, we assume that the number of source observations is the same as the channel uses. When these are not the same, the results can be deduced by considering a normalization coefficient. All the keys and random variables take values from some finite sets. Now, we state the conditions that should be met in the described secret key sharing framework as shown in Fig.1.

\begin{defi}
In the secret key agreement strategy of the proposed model, the rate pair $(R_{1},R_{2})$ is an achievable key rate pair if for every $\varepsilon>0$ and sufficiently large $N$, we have:
\begin{eqnarray}
&{\Pr \{(K_{1},K_{2}) \ne(\hat{K}_{1},\hat{K}_{2})\} <\varepsilon}\label {rel}
\\ &{{\tfrac{1}{N}} I(K_{1} ;S_{2}^{N} ,X_{2}^{N} ,Y_{2}^{N} )<\varepsilon }\label {sec1}
\\ &{{\tfrac{1}{N}} I(K_{2} ;S_{1}^{N} ,X_{1}^{N} ,Y_{1}^{N} )<\varepsilon } \label {sec2}
\\ &{{\tfrac{1}{N}} H(K_{1} )>R_{1} -\varepsilon {\rm \; and\; }{\tfrac{1}{N}} H(K_{2} )>R_{2} -\varepsilon }\label {ratdef}
\end{eqnarray}

\vspace{-.2cm}
\noindent Equation \eqref{rel} means that user 3 correctly estimates the secret keys and equations \eqref{sec1} and \eqref{sec2} mean that users 1 and 2 effectively have no information about each other's secret keys.
\end{defi}
\begin{defi}\label{defi:defreg}
The region containing all achievable secret key rate pairs $(R_{1},R_{2})$ is the secret key capacity region.
\end{defi}


\section{ Main Results }
\noindent Now, we state the main result of the paper.

\begin{theorem}
In the described model, all rate pairs in the closure of the convex hall of the set of all key rate pairs $(R_{1} ,R_{2} )$ that satisfy the following region are achievable:
\vspace{+.2cm}
\small{
\noindent $\left\{\begin{array}{l} {\!\!\!\!\!R_{1}>0,R_{2}>0}
\\[.12cm]{\!\!\!\!\!R_{1}\!\!\le\!\![I(\!U_{1};S_{3}|U_{2}\!)\!\!-\!\!I\!(\!U_{1};S_{2}|U_{2}\!)]^{+}\!\!+\!\![I\!(\!V_{1};Y_{3}|V_{2}\!)\!\!-\!\!I\!(\!V_{1};Y_{2}|V_{2},X_{2}\!)]^{+},}
\\[.12cm]{\!\!\!\!\!R_{2}\!\!\le\!\![I(\!U_{2};S_{3}|U_{1}\!)\!\!-\!\!I\!(\!U_{2};S_{1}|U_{1}\!)]^{+}\!\!+\!\![I\!(\!V_{2};Y_{3}|V_{1}\!)\!\!-\!\!I\!(\!V_{2};Y_{1}|V_{1},X_{1}\!)]^{+},}
\\[.12cm]{\!\!\!\!\!R_{1}\!+\!R_{2}\!\!\le\!\![I(\!U_{1},U_{2};S_{3}\!)\!\!-\!\!I\!(\!U_{1};S_{2}|U_{2}\!)\!\!-\!\!I\!(\!U_{2};S_{1}|U_{1}\!)\!\!-\!\!I\!(\!U_{1};U_{2}\!)]^{+}}
\\[.12cm]{\!\!\!\!\!+[I(\!V_{1},V_{2};Y_{3}\!)\!\!-\!\!I\!(\!V_{1};Y_{2}|V_{2},X_{2})\!\!-\!\!I\!(\!V_{2};Y_{1} |V_{1} ,X_{1})]^{+}}
\end{array}\right. $
}

\normalsize{
\noindent subject to the constraints:
\vspace{+.15cm}

}
\small{
\noindent $\begin{array}{l} {I(U_{1};S_{1}|U_{2},S_{3})\le I(V_{1};Y_{3}|V_{2}),I(U_{2};S_{2}|U_{1},S_{3})\le I(V_{2};Y_{3}|V_{1}),}
\\[.1cm]{I(U_{1},U_{2};S_{1},S_{2}|S_{3})\le I(V_{1},V_{2};Y_{3}),} \end{array}$
}

\vspace{-.05cm}
\normalsize{
\noindent for random variables taking values in sufficiently large finite sets according to the distributions:
}

\vspace{+.1cm}
\small{
\noindent $\begin{array}{l} {p_{1} (u_{1} ,u_{2} ,s_{1} ,s_{2} ,s_{3} )=p(u_{1} \left|s_{1} \right. )p(u_{2} \left|s_{2} \right. )p(s_{1} ,s_{2} ,s_{3} ),}
\\[.05cm]{p_{1}\!(\!v_{1}\!,\!v_{2},\!x_{1}\!,\!x_{2},\!y_{1}\!,\!y_{2},\!y_{3}\!)\!\!=\!\!p(\!v_{1}\!)p(\!v_{2}\!)p(\!x_{1}\!|v_{1}\!)p(\!x_{2}\!|v_{2}\!)p(\!y_{1},\!y_{2},\!y_{3} \!|x_{1},\!x_{2}\!).}
\end{array}$
}
\end{theorem}

\vspace{+.2cm}
\normalsize{
The proof of Theorem 1 is given in Appendix I. Here, the sketch of the proof is given. To derive this inner bound, a part of the keys between the users is generated by using the source common randomness and the other part is generated by exploiting the channel common randomness.  For this purpose, we impose the separation strategy as in \cite{joint-source-channel} to the GDMMAC.
\begin{figure}
\centering
\includegraphics[width=7cm,height=3.7cm]{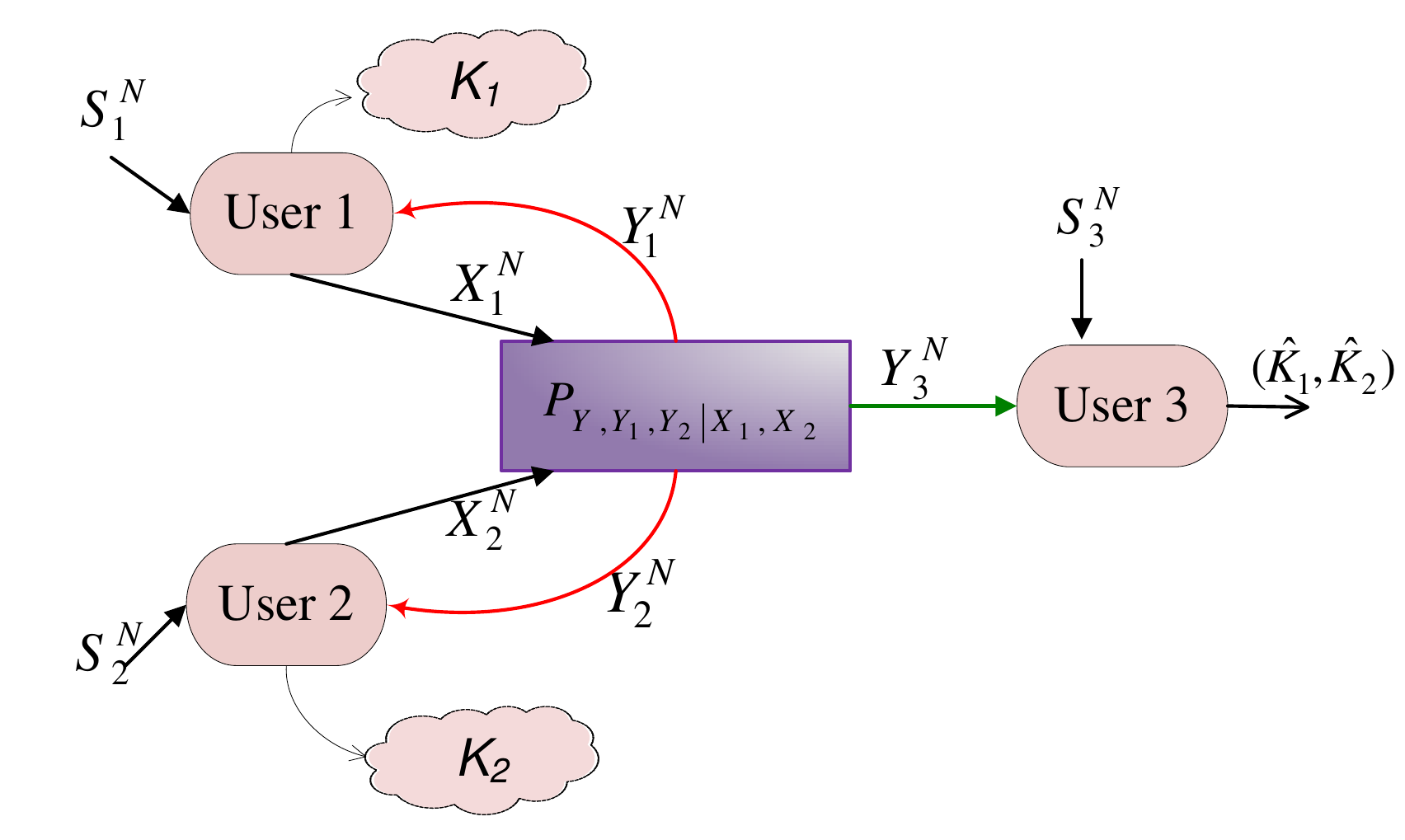}

\caption{\footnotesize{Secret key sharing using correlated sources over GDMMAC}
}
\vspace{-.7cm}
\label{fig_sim}
\end{figure}
At the first part of secret key sharing, users 1 and 2 generate secret keys using source observations. To share these keys with user 3, some information should be sent by users 1 and 2 via the GDMMAC according to Wyner-Ziv codebook for multiple sources as in \cite{Wyner-Ziv} and this information should satisfy the constraints of Theorem 1. The information sent by each user is independent of his secret key. On the other hand, the GDMMAC is noisy and not all of the information sent by user 1 can be decoded by user 2 and vice versa. Hence, part of the information sent via the GDMMAC can be itself used for secret key generation exploiting the channel common randomness. The latter rates are achievable if they belong to the secrecy rate region of the GDMMAC in \cite{mac-poor}.

\begin{pro}
The following region is an explicit outer bound of the secret key capacity region in the described setup:

\vspace{.1cm}
\noindent $\begin{array}{l} {0\le R_{1} \le I(X_{1} ;Y_{3} \left|X_{2} ,Y_{2} )\right. +I(S_{1} ;S_{3} \left|S_{2})\right.,}
\\{0\le R_{2} \le I(X_{2} ;Y_{3} \left|X_{1} ,Y_{1} )\right. +I(S_{2} ;S_{3} \left|S_{1} )\right.,}
\\{0\le R_{1}+R_{2} \le I(X_{1},X_{2} ;Y_{3})+I(S_{1},S_{2} ;S_{3}).}

\end{array}$
\end {pro}

\vspace{.15cm}
The proof of Proposition 1 is given in Appendix II.

It should be noted that the separation strategy used in achievability may not be optimal in general and hence, we could not yet derive the secret key capacity region. In the next section, we investigate a special case where the rate region in Theorem 1 can achieve the capacity region.

\section{Special Case}
\noindent In this section, we consider a special case of the general problem discussed in the previous sections. This special case arises when the channel inputs and outputs and the sources form Markov chains as:

\noindent $\begin{array}{l}
{(\!X_{1},X_{2}\!)\!-\!Y_{1}\!-\!Y_{3},{\rm \; \; \; \; \; \; \;}(\!X_{1},X_{2}\!)\!-\!Y_{2}\!-\!Y_{3},{\rm \; \; \; \; \; \; \; }S_{1}\!-\!S_{3}\!-\!S_{2}.}\end{array}$

\noindent Also, we constrain a special condition to the channel such that each of the channel inputs can be specified with access to the receiver's channel output and the other transmitter's input or in other words:

\noindent $\begin{array}{l}
{H(X_{1} \left|X_{2} \right. ,Y_{3} )=0,{\rm \; \; \; \; \; \; \; \; \; }H(X_{2} \left|X_{1} \right. ,Y_{3} )=0.}\end{array}$

In this case, the GDMMAC cannot itself provide secrecy; due to the Markov chains between the inputs and outputs, however, it can be used by users 1 and 2 to transmit the required information of the source common randomness. The secret key capacity region is given in the following theorem.

\begin{theorem}
In the mentioned special case of the described setup, the secret key capacity region is the set of all rate pairs $(R_{1},R_{2})$ that satisfy:
\vspace{-.1cm}
\[R_{1} \le I(U_{1} ;S_{3} \left|S_{2} )\right. ,{\rm \; \; \; \; \; }R_{2} \le I(U_{2} ;S_{3} \left|S_{1} )\right. ,\]

\vspace{-.25cm}
\noindent subject to the constraints:

\vspace{-.4cm}
\[\begin{array}{l} {I(U_{1};S_{1}|S_{3})\le I(X_{1};Y_{3}|X_{2}),I(U_{2};S_{2}|S_{3})\le I(X_{2};Y_{3}|X_{1}),}
\\{I(U_{1};S_{1}|S_{3})+I(U_{2};S_{2}|S_{3})\le I(X_{1},X_{2};Y_{3}),} \end{array}\]
 for random variables taking values in sufficiently large finite sets according to the distributions:
 \vspace{-.2cm}
\[\begin{array}{l}{p(u_{1},u_{2},s_{1},s_{2},s_{3})=p(u_{1}|s_{1})p(u_{2}|s_{2})p(s_{1},s_{2},s_{3}),}
\\{p_{1}(x_{1},x_{2},y_{1},y_{2},y_{3})=p(x_{1})p(x_{2})p(y_{1},y_{2},y_{3}|x_{1},x_{2} ).} \end{array}\]
\end{theorem}

The achievability can be deduced from Theorem 1 by substituting $V_{1}=X_{1},V_{2}=X_{2}$ and employing the Markov chains. The outer bound is proved in Appendix III. In this example, the GDMMAC is used just to transmit the required source common randomness information and hence, the problem can be treated as a pure source model with rate limited public channels. When the channel rates satisfy:
\vspace{+.1cm}

\noindent $\begin{array}{l}
{H(S_{1}|S_{3})\le I(X_{1};Y_{3}|X_{2}),H(S_{2}|S_{3})\le I(X_{2};Y_{3}|X_{1}),}\\{H(S_{1},S_{2}|S_{3})\le I(X_{1},X_{2};Y_{3}),}
\end{array}$
\vspace{+.1cm}

\noindent the case is similar to the special case of the forward source model of \cite{salimi-new-source-model} where the secret key capacity is the union of all rate pairs that satisfy:

\noindent $\begin{array}{l}
{R_{1} \le I(S_{1} ;S_{3} \left|S_{2} )\right. ,{\rm \; \; \; \; \; \; \; \; \; \; \; \; }R_{2} \le I(S_{2} ;S_{3} \left|S_{1} )\right. .}\end{array}$

\section{ A Gaussian Example}

\noindent As a scalar Gaussian example, we consider the case where the source observations at the users are according to Markov chain $S_{1}-S_{3}-S_{2}$. Then, without loss of generality, we can model them as ${S_{1}=S_{3}+E_{s,1},S_{2}=S_{3}+E_{s,2}}$ where $S_{3},E_{s,1}$ and $E_{s,2}$ are independent zero mean Gaussian variables with variances $P_{3},N_{s,1}$ and $N_{s,2}$, respectively. The GDMMAC is described by:

\vspace{+.2cm}
\small{
\noindent $\begin{array}{l}
{\!\!\!Y_{3}\!=\!X_{1}+X_{2}+E_{c,3},Y_{2}\!=\!X_{1}+X_{2}+E_{c,2},Y_{1}\!=\!X_{1}+X_{2}+E_{c,1},}
\end{array}$
}
\vspace{-.2cm}

\normalsize{
\noindent in which $E_{c,i}\!\sim\!{\rm{\mathcal N}}(0,N_{c,i})$ for $i\!=\!1,2,3$ with the input power constraints $P_{1}$ and $P_{2}$ at users 1 and 2, respectively. By the standard arguments as the discrete channel arguments, the results can be extended to Gaussian case. We consider:

$V_{1}=X_{1},V_{2}=X_{2},S_{1}=U_{1}+D_{1},S_{2}=U_{2}+D_{2}$
\vspace{+.1cm}

\noindent in Theorem 1 where \small{$U_{i}\!\sim\!{\rm{\mathcal N}}(0,P_{U_{i}}\!),D_{i}\!\sim\!{\rm{\mathcal N}}(0,P_{3}\!+\!N_{s,i}\!-\!P_{U_{i}}\!)$} and \normalsize{$U_{i}$ is independent of $D_{i}$ for $i=1,2$.}

\noindent By substituting the random variables as above in Theorem 1, the following proposition for secret key rate region can be deduced for which the proof is relinquished due to space limitation.
}

\begin{pro}
$(R_{1},R_{2})$ is an achievable key rate pair if is satisfies the equations at the top of the next page.

\begin{figure*}[!t]
\normalsize{
$\begin{array}{l} {R_{1}>0,R_{2}>0,}
\\ {R_{1}\le\frac{1}{2}\log(1+\frac{P_{U_{1}}P_{3}N_{s,2}}{((P_{3}+N_{s,1})^{2}-P_{U_{1}}P_{3})(P_{3}+N_{s,2})})+\frac{1}{2} [\log(1+\frac{P_{1}}{N_{c,3}})-\log(1+\frac{P_{1}}{N_{c,2}})]^{+},}
\\[.3cm]{R_{2}\le\frac{1}{2}\log(1+\frac{P_{U_{2}}P_{3}N_{s,1}}{((P_{3}+N_{s,2})^{2}-P_{U_{2}}P_{3})(P_{3}+\!N_{s,1})})+\frac{1}{2} [\log(1+\frac{P_{2}}{N_{c,3})})-\log(1+\frac{P_{2}}{N_{c,1}})]^{+},}
\\[.3cm]{R_{1}+R_{2}\le\frac{1}{2}\log[(1+\!\frac{P_{U_{1}}P_{3}N_{s,2}}{((P_{3}+N_{s,1})^{2}-P_{U_{1}}P_{3})(P_{3} +N_{s,2})})(1+\frac{P_{U_{2}}P_{3}N_{s,1}}{((P_{3}+N_{s,2})^{2}-P_{U_{2}}P_{3})(P_{3}+N_{s,1})})]}
\\[.3cm]{+\frac{1}{2} [\log (1+\frac{P_{1} +P_{2} }{N_{c,3} )} )-\log (1+\frac{P_{1} }{N_{c,2} } )-\log (1+\frac{P_{2} }{N_{c,1}})]^{+}}
\end{array}$
\vspace{+.2cm}

\noindent due to the constraints:

\vspace{+.2cm}
\noindent $\begin{array}{l} {P_{U_{1} } \le \frac{P_{1}(P_{3} +N_{s,1} )^{2} }{N_{s,1} N_{c,3} +P_{1}(P_{3} +N_{s,1} )} {\rm ,\; \; \; \; \; \; \; \; }P_{U_{2} } \le \frac{P_{2}(P_{3} +N_{s,2} )^{2} }{N_{s,2} N_{c,3} +P_{2}(P_{3} +N_{s,2} )} ,}
\\[.3cm]{(A_{1} -P_{U_{1} } )(A_{2} -P_{U_{2} } )\ge A_{1} A_{2} -\frac{(P_{1} +P_{2} )(P_{3} +N_{s,1} )^{2} (P_{3} +N_{s,2} )^{2} }{A} {\rm ,}}
\\[.2cm]{A=N_{s,1}N_{s,2}N_{c,3}+P_{3}N_{c,3}(N_{s,1}+N_{s,2})+(P_{1}+P_{2})(P_{3}+N_{s,1})(P_{3}+N_{s,2}),}
\\[.2cm]{A_{1} =\frac{(P_{3} +N_{s,1} )^{2} (N_{s,2} N_{c,3} +(P_{1} +P_{2} )(P_{3} +N_{s,2} ))}{A},A_{2} =\frac{(P_{3} +N_{s,2} )^{2} (N_{s,1} N_{c,3} +(P_{1} +P_{2} )(P_{3} +N_{s,1} ))}{A}.}
\end{array}$
}

\vspace{.1cm}
\hrulefill
\end{figure*}
\end{pro}

 For the values $P_{1}=P_{2}=P_{3}=1,N_{s,1}=N_{s,2}=N_{c,3}=0.5,$ the rate region is shown in Fig. 2 where  $N_{c,1}=N_{c,2}$ and they vary from 0.5 to 0.9. When $N_{c,1}=N_{c,2}=0.5$, no secret key can be generated  through GDMMAC and the second terms of the secret key rates bound in Proposition 2 will be zero. In this case, the region is the secret key rate region of a pure source model with rate limited public channels. When these noises increase, the key rate region enlarges however the sum rate boundary remains fixed since the sum rate term related to channel common randomness is zero until these noises amount to about $N_{c,1}=N_{c,2}=0.75$. At this point, the term related to the channel common randomness in the sum rate bound would be positive and hence, the sum rate bound increases. For $N_{c,1}=N_{c,2}=0.8$, the rate region is rectangular since each user's rate bound in proposition 2 will exceed the total sum rate bound. For $N_{c,1}=N_{c,2}=0.9$, this region becomes pentagonal as the sum rate due to channel common randomness is increased such that the total sum rate bound dominates each user's rate bound.}

\begin{figure}
\centering
\vspace{-.5cm}
\includegraphics[width=8cm,height=5.5cm]{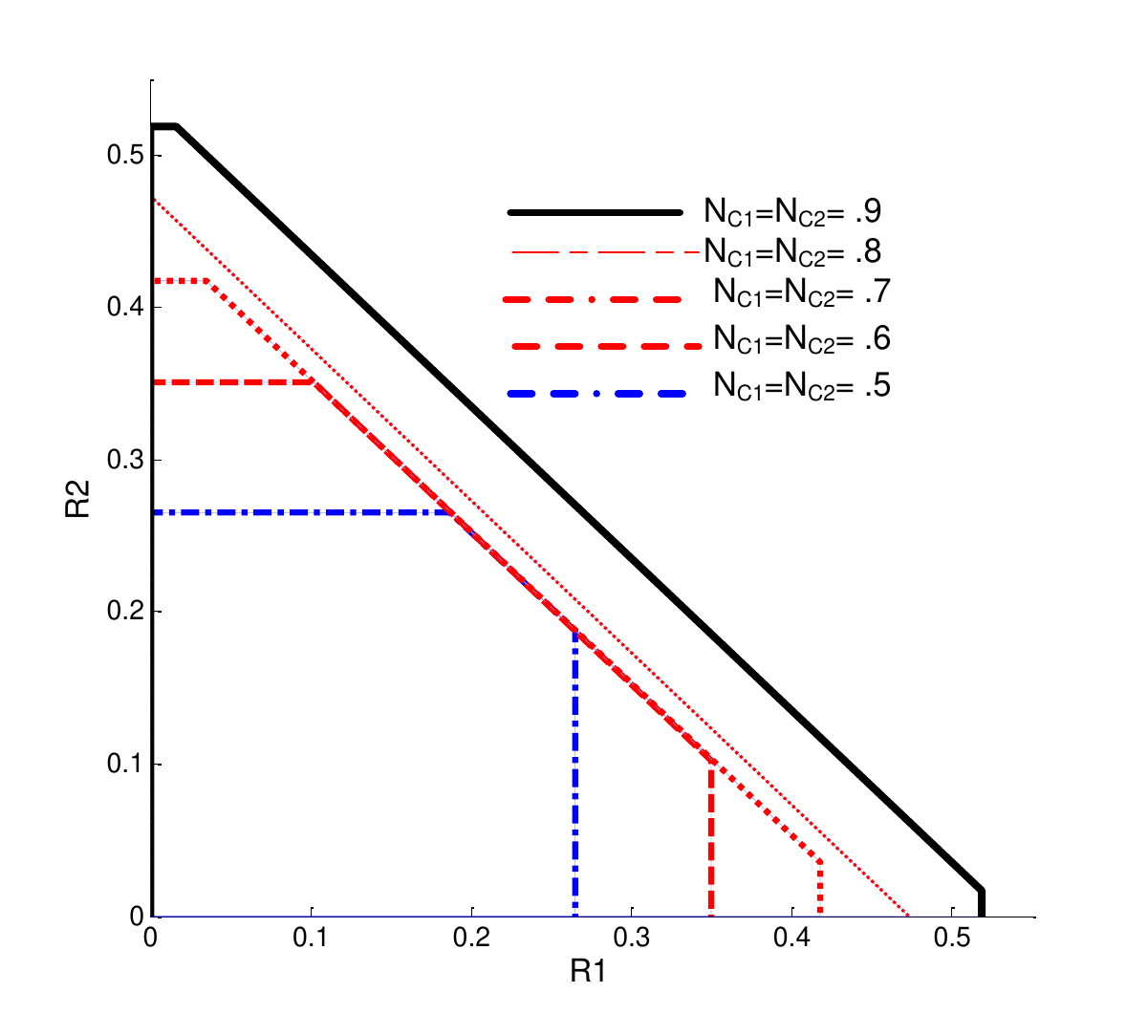}
\vspace{-.5cm}
\caption{\footnotesize{Secret key rate region for different values of $N_{c1}$ and $N_{c2}$}
}
\vspace{-.6cm}
\label{fig_sim}
\end{figure}

\section{Conclusion}
\noindent The problem of secret key sharing in a combined framework with both source and the channel common randomness was studied. For this problem, the inner bound and the explicit outer bound of the secret key capacity region were derived. The separation strategy used to achieve the inner bound was not optimal in general, however, for the especial case where the channel inputs and outputs and the sources form Markov chains in some order, the secret key capacity region was derived. It was shown that this case can be simulated as the source model of key agreement with public channels of limited capacity. As future work, the joint source channel coding can be investigated instead of the separation strategy. In addition, in a more practical scenario, some types of dependency between the source and the channel can be supposed.

\section*{Appendix I}
\subsection*{Proof of Theorem 1}
\noindent We fix the distribution to be the same as in Theorem 1. Users 1 and 2 independently and randomly generate typical sequences $u_{1}^{N}$ and $u_{2}^{N}$, respectively, each with probability:
\vspace{-.2cm}
\begin{equation}
p(u_{1}^{N} )=\prod _{i=1}^{N}p(u_{1,i} ) ,p(u_{2}^{N} )=\prod _{i=1}^{N}p(u_{2,i} ) .\no
\end{equation}
\vspace{-.3cm}

\noindent The number of the sequences $u_{1}^{N}$ and $u_{2}^{N}$ are $2^{N(I(U_{1};S_{1})+\varepsilon')}$, and $2^{N(I(U_{2};S_{2})+\varepsilon')}$, respectively, in which $\varepsilon'\!\!>\!0$ can be chosen arbitrarily small. Using two-layered random binning, they are labeled as:
\vspace{+.1cm}

\small{
\noindent $\begin{array}{l}
{u_{1}^{N}(k_{1S},k'_{1S},k''_{1S}),}
\\{k_{1S}\in\{1,...,2^{Nr_{1S})}\},k'_{1S}\in\{1,...,2^{Nr'_{1S}}\} ,k''_{1S}\in\{1,...,2^{Nr''_{1S}}\}}
\\{u_{2}^{N}(k_{2S},k'_{2S},k''_{2S}),}
\\{k_{2S}\in\{1,...,2^{Nr_{2S})}\},k'_{2S}\in\{1,...,2^{Nr'_{2S}}\},k''_{2S}\in\{1,...,2^{Nr''_{2S}}\}}
\end{array}$
}

\noindent where:
\vspace{-.2cm}
\begin{eqnarray}{\!\!\!\!r_{1S}\!+\!r'_{1S}\!=\!I(U_{1};S_{1})\!-\!I(U_{1};S_{2})\!+\!2\varepsilon',r''_{1S}\!=\!I(U_{1};S_{2})\!-\!\varepsilon',}\label{ratedef1} \\{\!\!\!\!r_{2S}\!+\!r'_{2S}\!=\!I(U_{2};S_{2})\!-\!I(U_{2};S_{1})\!+\!2\varepsilon',r''_{2S}\!=\!I(U_{2};S_{1})\!-\!\varepsilon',}\label{ratedef2}
\end{eqnarray}
\normalsize{
It is obvious that $r_{1S}+r'_{1S}+r''_{1S}=I(U_{1};S_{1})+\varepsilon'$ and so, each sequence $u_{1}^{N}$ can be determined if the indices $(k_{1S},k'_{1S},k''_{1S})$ are known. The same is true for $u_{2}^{N}$.
}

 In addition to the secret key codebooks, the wiretap channel codebooks are generated. Users 1 and 2 generate independent sequences $v_{1}^{N}$ and $v_{2}^{N}$, respectively, each with probability:
\vspace{-.2cm}

\begin{equation}
p(v_{1}^{N} )=\prod _{i=1}^{N}p(v_{1,i} ) ,p(v_{2}^{N} )=\prod _{i=1}^{N}p(v_{2,i} )\no
\end{equation}
\vspace{-.2cm}

\noindent The number of sequences $v_{1}^{N}$ and $v_{2}^{N}$ is $2^{N(r_{1C}+r'_{1C})}$ and $2^{N(r_{2C}+r'_{2C})}$, respectively, which are labeled as:

$\begin{array}{l} {v_{1}^{N}(k_{1C},k'_{1C}),k_{1C}\in\{ 1,...,2^{Nr_{1C}}\},k'_{1C}\in\{1,...,2^{Nr'_{1C}}\}}
\\{v_{2}^{N}(k_{2C},k'_{2C}),k_{2C}\in\{1,...,2^{Nr_{2C}}\},k'_{2C}\in\{1,...,2^{Nr'_{2C}}\}}
\end{array}$

\vspace{.1cm}
\noindent where:
\vspace{-.1cm}
\small{
\begin{equation}
r'_{1C}=I(V_{1};X_{2},Y_{2}|V_{2})-\varepsilon'',r'_{2C}=I(V_{2};X_{1},Y_{1}|V_{1})-\varepsilon''\label{ratedef3}
\end{equation}}
\vspace{-.2cm}
\normalsize{
Two functions $f_{1}$ and $f_{2}$ are defined as:}
\vspace{+.1cm}
\[\begin{array}{l} {f_{1} :{\rm {\mathcal V}}_{1} \to {\rm {\mathcal K}}'_{1S} ,{\rm \; \; \; \; \; \; \; \; \; \; \; \; }f_{2} :{\rm {\mathcal V}}_{2} \to {\rm {\mathcal K}}'_{2S} ,{\rm \; \; \; \; \; \; \; \; }} \\ {{\rm {\mathcal K}}'_{1S} =\{ 1,...,2^{Nr'_{1S} } \} ,{\rm \; \; \; \; }{\rm {\mathcal K}}'_{2S} =\{ 1,...,2^{Nr'_{2S} } \} ,} \end{array}\]

\noindent where ${\rm{\mathcal V}}_{1}$ and ${\rm{\mathcal V}}_{2}$ are the set of $2^{N(r_{1C}+r'_{1C})}$ and $2^{N(r_{2C}+r'_{2C})}$ codewords $v_{1}^{N}$ and $v_{2}^{N}$, respectively. Mapping $f_{1}$ is a random partitioning of codewords $v_{1}^{N}$ into $2^{Nr'_{1S}}$ equal-sized parts. Elements of part $i$ are labeled as $({\rm{\mathcal V}}_{1})_{i}$. Mapping $f_{2}$ is similarly defined. In the definition of $f_{1}$, we assume that $r_{1C}+r'_{1C}\!\ge\!r'_{1S}$ and as we would see in the decoding step, this can be deduced from the constraint of Theorem 1. The same is true for $f_{2}$.

For encoding, when typical sequences $s_{1}^{N}$ and $s_{2}^{N}$ are observed at users 1 and 2, respectively, sequences $u_{1}^{N}$ and $u_{2}^{N}$ are chosen at the corresponding users such that $(s_{1}^{N},u_{1}^{N})$ and $(s_{2}^{N},u_{2}^{N})$ are $\varepsilon'-$jointly typical. It can be seen that these sequences are unique with high probability for arbitrarily small $\varepsilon'$. For the sequences $u_{1}^{N}(k_{1S},k'_{1S},k''_{1S})$ and $u_{2}^{N}(k_{2S} ,k'_{2S},k''_{2S})$, the indices $k_{1S}$ and $k_{2S}$ are chosen by users 1 and 2, respectively, to share with user 3 as the secret keys due to the source observations. For this purpose, the indices $k'_{1S} $ and $k'_{2S} $ are the required information to be sent by users 1 and 2 to user 3.  User 1 encodes $k'_{1S}$ in such a way that he returns $v_{1}^{N}$ randomly chosen from $({\rm{\mathcal V}}_{1})_{k'_{2S}}$ using the mapping $f_{1}$. User 2 acts in a same way using mapping $f_{2}$ and returns $v_{2}^{N}$. For the selected $v_{1}^{N}(k_{1C},k'_{1C})$ and $v_{2}^{N}(k_{2C},k'_{2C})$, user 1 and 2, respectively, consider $k_{1C}$ and $k_{2C}$ to share with user 3 as the secret keys due to the channel common randomness. Then, the channel inputs  $x_{1}^{N}$ and $x_{2}^{N}$ are sent over GDMMAC according to the distributions $p(x_{1}|v_{1})$ and $p(x_{2}|v_{2})$.

For decoding, user 3, first, chooses the sequences $v_{1}^{N}$ and $v_{2}^{N}$ which are $\varepsilon_{1}-$jointly typical with the received $y_{3}^{N}$ where $\varepsilon_{1}=\frac{\varepsilon}{8}$. User 3 decodes key pair $(k_{1C},k_{2C})$ if $(v_{1}^{N}(k_{1C},k'_{1C}),v_{2}^{N}(k_{2C},k'_{2C}),y_{3}^{N})\in A_{\varepsilon_{1}}^{(N)}(P_{V_{1},V_{2},Y_{3}}),$ when such $(v_{1}^{N}(k_{1C},k'_{1C}),v_{2}^{N}(k_{2C},k'_{2C}))$ exists and is unique. Otherwise, it declares error. It can be shown that the decoding error probability of this step is bounded as:
\vspace{-.05cm}
\[\begin{array}{l} {P_{e1}^{(N)}\le \varepsilon_{1}+2^{N(r_{1C}+r'_{1C}+r_{2C}+r'_{2C}-(I(V_{1},V_{2};Y_{3})-4\varepsilon_{1}))+}}
\\{2^{N(r_{1C}+r'_{1C}\!-(I(V_{1};Y_{3}|V_{2})\!-3\varepsilon_{1}))}\!+\!2^{N(r_{2C}+r'_{2C}\!-\!(I(V_{2};Y_{3}|V_{1})\!-3\varepsilon_{1}))} .}
\end{array}\]
\vspace{-.35cm}

If we set:
\vspace{+.2cm}

\small{
\noindent $\begin{array}{l} {\!\!\!r_{1C}\!<\!I\!(\!V_{1};\!Y_{3}|V_{2}\!)\!-\!I\!(\!V_{1};\!X_{2},\!Y_{2}|V_{2}\!),\!r_{2C}\!<\!I\!(\!V_{2};\!Y_{3}|V_{1}\!)\!-\!I\!(\!V_{2};\!X_{1},\!Y_{1}|V_{1}\!),}
\\{r_{1C}+r_{2C} <I(V_{1},V_{2};Y_{3})-I(V_{1};X_{2},Y_{2}|V_{2})-I(V_{2};X_{1},Y_{1}|V_{1}),}
\end{array}$

\vspace{+.2cm}
\normalsize{
\noindent then by substituting $r'_{1C}$ and $r'_{2C}$ from \eqref{ratedef3}, we have:
\vspace{+.2cm}
}

\noindent $P_{e1}^{(N)} \le \varepsilon _{1} +2^{N(-2\varepsilon ''+4\varepsilon _{1} )} +2^{N(-\varepsilon ''+3\varepsilon _{1} )} +2^{N(-\varepsilon ''+3\varepsilon _{1} )} .$

\noindent By setting $\varepsilon''\!>\!3\varepsilon_{1}\!=\!\frac{3\varepsilon}{8}$, for example $\varepsilon''\!=\!4\varepsilon_{1}\!=\!\frac{\varepsilon}{2}$, we choose $N$ so large that $2^{N(-\varepsilon''\!+\!3\varepsilon_{1})}\!\le\!\varepsilon_{1}$ , and then $P_{e1}^{(N)}\le 4\varepsilon_{1} \!=\!\frac{\varepsilon}{2}$.

\noindent After correct decoding $(k_{1C},k'_{1C})$ and $(k_{2C},k'_{2C})$, user 3 considers $k_{1C}$ and $k_{2C}$ to share with users 1 and 2, respectively, as the secret keys due to the channel common randomness. Then, using functions $f_{1}$ and $f_{2}$, user 3 finds the mappings $({\rm{\mathcal V}}_{1})_{i}$ of the codeword $v_{1}^{N}$ and $({\rm {\mathcal V}}_{2})_{j}$ of the codeword $v_{2}^{N}$ and sets $k'_{1S}=i,k'_{2S}=j$. Now, user 3 decodes sequences $u_{1}^{N}(k_{1S},k'_{1S},k''_{1S})$ and $u_{2}^{N}(k_{2S},k'_{2S},k''_{2S})$ such that:

\vspace{+.2cm}
\small{
\noindent
\[(u_{1}^{N}(\!k_{1S},k'_{1S}\!=\!i,k''_{1S}\!),u_{2}^{N}(\!k_{2S},k'_{2S}\!=\!j,k''_{2S}\!),s_{3}^{N})\!\in\!\!A_{\varepsilon _{2}}^{(N)}(\!P_{U_{1},U_{2},S_{3}}\!),\]}
\vspace{-.4cm}

\normalsize{
\noindent when such $(u_{1}^{N}(k_{1S},k'_{1S},k''_{1S}),u_{2}^{N}(k_{2S},k'_{2S},k''_{2S}))$ exists and is unique. Otherwise, it declares error. For $\varepsilon_{2}=\frac{\varepsilon}{16}$, according to Wyner-Ziv problem for multiple sources in \cite{Wyner-Ziv}, the decoding error probability of this step is bounded as:
}
\[\begin{array}{l} {P_{e2}^{(N)} \le \varepsilon _{2} +2^{N(14\varepsilon _{2} +I(U_{1} ,U_{2} ;S_{1} ,S_{2} \left|S_{3} )\right. -(r'_{1S} +r'_{2S} ))} }
\\ {+2^{N(7\varepsilon _{2} +I(U_{1} ;S_{1} \left|U_{2} ,S_{3} )\right. -r'_{1S} )} +2^{N(7\varepsilon _{2} +I(U_{2} ;S_{2} \left|U_{1} ,S_{3} )\right. -r'_{2S} )} .} \end{array}\]
\vspace{-.1cm}
If we set:
\small{
\[\begin{array}{l} {\!\!r_{1S}<I(U_{1};S_{3}|U_{2})-I(U_{1};S_{2}|U_{2}),}
\\{\!\!r_{2S}<I(U_{2};S_{3}|U_{1})-I(U_{2};S_{1}|U_{1}),}
\\{\!\!r_{1S}\!+\!r_{2S}\!<\!I(U_{1},\!U_{2};\!S_{3}\!)\!-\!I\!(U_{1};\!S_{2}|U_{2}\!)\!-\!I\!(U_{2};\!S_{1}|U_{1}\!)\!-\!I\!(U_{1};U_{2}\!),}\end{array}\]
}
\vspace{-.1cm}

\normalsize{
\noindent then by substituting $r'_{1S}$ and $r'_{2S}$ from \eqref{ratedef1} and \eqref{ratedef2}, we have:
}
\vspace{-.1cm}
\[P_{e2}^{(N)} \le \varepsilon _{2} +2^{N(14\varepsilon _{2} -4\varepsilon ')} +2^{N(7\varepsilon _{2} -2\varepsilon ')} +2^{N(7\varepsilon _{2} -2\varepsilon ')} .\]
\vspace{-.5cm}

\noindent By setting $\varepsilon'>\frac{7}{2}\varepsilon _{2}$, for example $\varepsilon'=4\varepsilon_{2}=\frac{\varepsilon}{4}$, we choose $N$ so large that $2^{N(14\varepsilon_{2}-4\varepsilon')}\le\varepsilon_{2}$, and then $P_{e2}^{(N)}\le 4\varepsilon _{2}=\frac{\varepsilon}{4}$. After this step of decoding, user 3 considers $k_{1S}$ and $k_{2S}$ to share with users 1 and 2, respectively, as the secret keys due to the source observations.

By the above arguments, the total decoding error probability is bounded as $P_{e}^{(N)}\le P_{e1}^{(N)}+P_{e2}^{(N)}\le \frac{\varepsilon}{2}+\frac{\varepsilon}{4}<\varepsilon$.

For correct decoding in the first step, according to $P_{e1}^{(N)}$ equation, it is necessary that:
\[\begin{array}{l} {r_{1C} {\rm +}r'_{1C} <I(V_{1} ;Y_{3} \left|V_{2} )\right. ,r_{2C} {\rm +}r'_{2C} <I(V_{2} ;Y_{3} \left|V_{1} )\right. } \\ {r_{1C} {\rm +}r'_{1C} +r_{2C} {\rm +}r'_{2C} <I(V_{1} ,V_{2} ;Y_{3} )} \end{array}\]
and hence, considering the rate constraints in Theorem 1, the functions $f_{1}$ and $f_{2}$ can be defined.

Now, we should check the security conditions of definition 1 and show that the indices $(k_{1S} ,k_{1C} )$ can be shared as secret keys between user 1 and user 3 and $(k_{2S},k_{2C})$ can be shared as secret keys between user 2 and user 3. We give the proof of \eqref{sec1} and by symmetry, \eqref{sec2} can be deduced. We have:
\begin{eqnarray}
&{I(K_{1S} ,K_{1C} ;S_{2}^{N} ,X_{2}^{N} ,Y_{2}^{N} )=}\no
\\&{\underbrace{I(K_{1S} ;S_{2}^{N} ,X_{2}^{N} ,Y_{2}^{N} \left|K_{1C} \right. )}_{A}+\underbrace{I(K_{1C} ;S_{2}^{N} ,X_{2}^{N} ,Y_{2}^{N} )}_{B}}\no
\end{eqnarray}
Now, we analyze each term separately. We consider $\varepsilon_{3}=\varepsilon_{4}=\varepsilon_{5}=\varepsilon_{6}=\frac{\varepsilon}{8}$. Some Markov chains useful in the security analysis are given in continue. These Markov chains arise from the coding scheme.
\small{
\begin{align}
& {\!\!\!(\!U_{1}^{N},\!U_{2}^{N}\!)\!-\!(\!K'_{1S},\!K'_{2S}\!)\!-\!(\!V_{1}^{N},\!V_{2}^{N}\!)\!-\!(\!X_{1}^{N},\!X_{2}^{N}\!)\!-\!(\!Y_{1}^{N},\!Y_{2}^{N},\!Y_{3}^{N}\!)}\label{markov1}
 \\& {U_{1}^{N} -S_{1}^{N} -S_{2}^{N} -U_{2}^{N}}\label{markov2}
 \\& {V_{1}^{N} -V_{2}^{N} -K'_{2S} -U_{2}^{N} -S_{2}^{N}}\label{markov3}
 \end{align}
 }
 \normalsize{
For term $A$, we have:}
\small{
\[\begin{array}{l} {I(K_{1S} ;S_{2}^{N} ,X_{2}^{N} ,Y_{2}^{N} \left|K_{1C} \right. )\le I(K_{1S} ;S_{2}^{N} ,X_{2}^{N} ,Y_{2}^{N} ,K_{1C} )}
\\ [.1cm]{\le I(K_{1S} ;S_{2}^{N} ,X_{2}^{N} ,Y_{2}^{N} ,K_{1C} ,K'_{1S} ,K'_{2S} )}
\\ {\mathop{=}\limits^{(a)} I(K_{1S} ;S_{2}^{N} ,K'_{1S} ,K'_{2S} )}
\\ {\mathop{=}\limits^{(b)} I(K_{1S} ;S_{2}^{N} ,K'_{1S} )=H(K_{1S} )-H(K_{1S} \left|S_{2}^{N} ,K'_{1S} )\right. }
\\ [.1cm]{=H(K_{1S} )-H(K_{1S} ,U_{1}^{N} \left|S_{2}^{N} ,K'_{1S} )\right. +H(U_{1}^{N} \left|S_{2}^{N} ,K'_{1S} ,K_{1S} )\right. }
\\ {\mathop{\le }\limits^{(c)}H(K_{1S})-H(K_{1S},U_{1}^{N}|S_{2}^{N} ,K'_{1S})+N\varepsilon_{3}}
\\ [.1cm]{\le H(K_{1S})-H(U_{1}^{N} |S_{2}^{N},K'_{1S})+N\varepsilon _{3} }
\\ [.1cm]{\le H(K_{1S} )-H(U_{1}^{N} ,K'_{1S} \left|S_{2}^{N} )\right. +H(K'_{1S} \left|S_{2}^{N} )\right. +N\varepsilon _{3} }
\\ {\mathop{=}\limits^{(d)} H(K_{1S} )-H(U_{1}^{N} \left|S_{2}^{N} )\right. +H(K'_{1S} \left|S_{2}^{N} )\right. +N\varepsilon _{3} }
\\ [.1cm]{\le H(K_{1S} )-H(U_{1}^{N} \left|S_{2}^{N} )\right. +H(K'_{1S} )+N\varepsilon _{3} }
\\ [.1cm]{=N(I(U_{1} ;S_{1} )-I(U_{1} ;S_{2} )+2\varepsilon ')-H(U_{1}^{N} \left|S_{2}^{N} )\right. +N\varepsilon _{3} }
\\ {\mathop{\le }\limits^{(e)} N(I(U_{1} ;S_{1} )\!-\!I(U_{1} ;S_{2} )\!+\!2\varepsilon ')\!-\!NH(U_{1}^{} \left|S_{2}^{} )\right. \!+N\varepsilon _{4} \!+N\varepsilon _{3} }
\\ [.1cm]{\le N\varepsilon _{3} +2N\varepsilon '+N\varepsilon _{4} =N\frac{3\varepsilon }{4} .} \end{array}\]}
\normalsize{
\noindent In above equations, (a) and (b) follow from Markov chains \eqref{markov1} and \eqref{markov2}, respectively. (c) can be deduced from the same approach as lemma 2 in ý[6] to show $H(U_{1}^{N}|S_{2}^{N},M'_{1},M_{1})\le N\varepsilon_{3}$. (d) follows from the fact that $k'_{1S}$ is one of the indices of $u_{1}^{N}$. To prove (e), the same approach as lemma 3 in ý[6] is exploited to show $NI(U_{1}|S_{2})\le I(U_{1}^{N}|S_{2}^{N})+N\varepsilon_{4}$.
}

For term $B$, we have:
\vspace{.15cm}

\small{
\noindent $\begin{array}{l}
{\!\!\!I(K_{1C} ;S_{2}^{N} ,X_{2}^{N} ,Y_{2}^{N} )\le I(K_{1C} ;S_{2}^{N} ,X_{2}^{N} ,Y_{2}^{N} ,V_{2}^{N} )}
\\[.1cm]{\!\!\!=H(K_{1C} )-H(K_{1C} \left|S_{2}^{N} ,X_{2}^{N} ,Y_{2}^{N} ,V_{2}^{N} \right. )}
\\[.1cm] {\!\!\!=H(K_{1C})-H(K_{1C},V_{1}^{N}|S_{2}^{N},X_{2}^{N},Y_{2}^{N},V_{2}^{N})+}
\\[.1cm] {{\ }H(V_{1}^{N} |K_{1C},S_{2}^{N},X_{2}^{N},Y_{2}^{N},V_{2}^{N})}
\\ {\!\!\!\mathop{\le }\limits^{(a)} H(K_{1C} )-H(K_{1C} ,V_{1}^{N} \left|S_{2}^{N} ,X_{2}^{N} ,Y_{2}^{N} ,V_{2}^{N} \right. )+N\varepsilon _{5} }
\\ {\!\!\!\mathop{=}\limits^{(b)} H(K_{1C} )-H(V_{1}^{N} \left|S_{2}^{N} ,X_{2}^{N} ,Y_{2}^{N} ,V_{2}^{N} \right. )+N\varepsilon _{5} }
\\ {\!\!\!\mathop{=}\limits^{(c)} H(K_{1C} )-H(V_{1}^{N} \left|X_{2}^{N} ,Y_{2}^{N} ,V_{2}^{N} \right. )+N\varepsilon _{5} }
\\ {\!\!\!\mathop{\le }\limits^{(d)} H(K_{1C} )-NH(V_{1} \left|X_{2}^{} ,Y_{2}^{} ,V_{2}^{} \right. )+N\varepsilon _{5} +N\varepsilon _{6} }
\\{\!\!\!\mathop{\le }\limits^{(e)}N(I(V_{1};Y_{3}|V_{2})-I(V_{1};X_{2},Y_{2}|V_{2}))-NH(V_{1} |X_{2}^{},Y_{2}^{},V_{2}^{})+}
\\[.1cm] {{\ \ }N\varepsilon_{5}+N\varepsilon_{6}}
\\[.1cm] {\!\!\!=-NH(V_{1} \left|Y_{3}^{} ,V_{2}^{} \right. )+N\varepsilon _{5} +N\varepsilon _{5} }
\\[.1cm] {\!\!\!\le N\varepsilon _{5} +N\varepsilon _{6} =N\frac{\varepsilon }{4} }
\end{array}$
}

\normalsize{
In the above equations, (a) can be deduced from the same approach as lemma 2 in ý[6] to show} \small{$H(V_{1}^{N}\!|K_{1C},\!X_{2}^{N},\!Y_{2}^{N}\!,V_{2}^{N}\!)\!\le\!N\!\varepsilon_{5}$}. \normalsize{(b) follows from the fact that $k_{1C}$ is one of the indices of $v_{1}^{N}$.(c) can be followed from Markov chain \eqref{markov3}. The same approach as lemma 3 in ý[6] can be exploited to show} \small{$NH(V_{1}\!|X_{2},\!Y_{2},V_{2\!})\!\!\le\!\! H(\!V_{1}^{N}\!|X_{2}^{N},\!Y_{2}^{N},\!V_{2}^{N}\!)\!\!+\!\!N\varepsilon_{6}$}. \normalsize{(e) is a direct consequence of reliable decoding at user 3.}

\noindent Hence, the security condition \eqref{sec1} is satisfied as:
\begin{equation}
I(K_{1S} ,K_{1C} ;S_{2}^{N} ,X_{2}^{N} ,Y_{2}^{N} )\le N(\frac{3\varepsilon }{4} +\frac{\varepsilon }{4} )=N\varepsilon.\no
\end{equation}

To show that the total rate of user 1's key is sum of $r_{1S}$ and $r_{1C}$, we should prove the independency of the keys of the two steps. When analyzing term $A$ of the security condition, we show that $I(\!K_{1S}\!;S_{2}^{N},\!X_{2}^{N},Y_{2}^{N},K_{1C})\!\le\!N\frac{3\varepsilon}{4}$ and hence $I(\!K_{1S};\!K_{1C}\!)\!\le\!N\varepsilon$, and this completes the proof of theorem 1.

\noindent
\section*{Appendix II}

\noindent
\subsection*{Proof of the explicit outer bound in Lemma 1}

\noindent First, users 1 and 2 generate secret keys $K_{1}$ and $K_{2}$, respectively, as stochastic functions $K_{1}=f_{1}(S_{1}^{N})$ and $K_{2}=f_{2}(S_{2}^{N})$. Then, they determine the channel inputs as stochastic functions $X_{1}^{N}=f_{3}(S_{1}^{N})$ and $X_{2}^{N}=f_{4}(S_{2}^{N})$ and send them via The GDMMAC. Consequently, $Y_{1}^{N},Y_{2}^{N}$ and $Y_{3}^{N}$ are received by user 1, user 2 and user 3, respectively. User 3 should be able to reconstruct the secret keys and according to Fano's inequality for an arbitrary small $\varepsilon>0$:
\begin{equation}
H(K_{1},K_{2}|Y_{3}^{N},S_{3}^{N})\le N(\frac{H(\varepsilon)}{N}+\varepsilon(\log |{\rm{\mathcal K}}_{{\rm 1}}||{\rm {\mathcal K}}_{{\rm 2}} |-1))\triangleq N\varepsilon_{1},\no
\end{equation}

\noindent where $\varepsilon _{1}\to0$ if $\varepsilon\to0$. Also, the following security conditions should be satisfied as:
\begin{equation}
I(K_{1} ;S_{2}^{N} ,X_{2}^{N} ,Y_{2}^{N} )<N\varepsilon {\rm ,\; \; \; \; \; }I(K_{2} ;S_{1}^{N} ,X_{1}^{N} ,Y_{1}^{N} )<N\varepsilon.\no
\end{equation}
Now, we show that for the secret keys satisfying the reliability and security conditions described above, we can deduce the explicit outer bound of Lemma 1. We prove the outer bound on $R_{1}$ (symmetry can be used for $R_{2}$).

\small{
\noindent $\begin{array}{l} {\!\!\!R_{1} =\frac{1}{N} H(K_{1} )\mathop{\le }\limits^{{\rm (a)}} \frac{1}{N} H(K_{1} \left|S_{2}^{N} ,X_{2}^{N} ,Y_{2}^{N} )\right. +\varepsilon }
\\ {\!\!\!\mathop{\le }\limits^{{\rm (b)}} \frac{1}{N} H(K_{1} \left|S_{2}^{N} ,X_{2}^{N} ,Y_{2}^{N} )\right. -\frac{1}{N} H(K_{1} \left|S_{3}^{N} ,Y_{3}^{N} )\right. +\varepsilon _{1} +\varepsilon }
\\ [.1cm]{\!\!\!\le \frac{1}{N} I(K_{1} ;S_{3}^{N} ,Y_{3}^{N} \left|S_{2}^{N} ,X_{2}^{N} ,Y_{2}^{N} )\right. +\varepsilon _{1} +\varepsilon }
\\ [.1cm]{\!\!\!\le \frac{1}{N} I(S_{1}^{N} ,X_{1}^{N} ;S_{3}^{N} ,Y_{3}^{N} \left|S_{2}^{N} ,X_{2}^{N} ,Y_{2}^{N} )\right. +\varepsilon _{1} +\varepsilon }
\\[.1cm] {\!\!\!=\frac{1}{N}I(S_{1}^{N},X_{1}^{N};S_{3}^{N}|S_{2}^{N},X_{2}^{N},Y_{2}^{N})+}
\\[.1cm]{{\ }\frac{1}{N}I(S_{1}^{N},X_{1}^{N};Y_{3}^{N}|S_{3}^{N},S_{2}^{N},X_{2}^{N},Y_{2}^{N})+\varepsilon_{1}+\varepsilon}
\\[.1cm] {\!\!\!=\frac{1}{N} (H(S_{3}^{N} \left|S_{2}^{N} ,X_{2}^{N} ,Y_{2}^{N} )\right. -H(S_{3}^{N} \left|S_{1}^{N} ,X_{1}^{N} ;S_{2}^{N} ,X_{2}^{N} ,Y_{2}^{N} )\right. )}
\\[.1cm] {\!\!\!+\frac{1}{N} I(S_{1}^{N} ,X_{1}^{N} ;Y_{3}^{N} \left|S_{3}^{N} ,S_{2}^{N} ,X_{2}^{N} ,Y_{2}^{N} )\right. +\varepsilon _{1} +\varepsilon }
\\ [.1cm]{\!\!\!\le \frac{1}{N} (H(S_{3}^{N} \left|S_{2}^{N} )\right. -H(S_{3}^{N} \left|S_{1}^{N} ,X_{1}^{N} ,S_{2}^{N} ,X_{2}^{N} ,Y_{2}^{N} )\right. )}
\\[.1cm]{\!\!\!+\frac{1}{N} I(S_{1}^{N} ,X_{1}^{N} ;Y_{3}^{N} \left|S_{3}^{N} ,S_{2}^{N} ,X_{2}^{N} ,Y_{2}^{N} )\right. +\varepsilon _{1} +\varepsilon }
\\ {\!\!\!\mathop{=}\limits^{{\rm (c)}} \frac{1}{N} (H(S_{3}^{N} \left|S_{2}^{N} )\right. -H(S_{3}^{N} \left|S_{1}^{N} ,S_{2}^{N} )\right. )}
\\[.1cm]{\!\!\!+\frac{1}{N} I(S_{1}^{N} ,X_{1}^{N} ;Y_{3}^{N} \left|S_{3}^{N} ,S_{2}^{N} ,X_{2}^{N} ,Y_{2}^{N} )\right. +\varepsilon _{1} +\varepsilon }
\\ [.1cm]{\!\!\!\le \frac{1}{N} (I(S_{3}^{N} ;S_{1}^{N} \left|S_{2}^{N} )\right. )+\frac{1}{N} (H(Y_{3}^{N} \left|X_{2}^{N} ,Y_{2}^{N} )\right. }
\\[.1cm]{\!\!\!-H(Y_{3}^{N} \left|S_{1}^{N} ,X_{1}^{N} ,S_{3}^{N} ,S_{2}^{N} ,X_{2}^{N} ,Y_{2}^{N} )\right. )+\varepsilon _{1} +\varepsilon }
\\ {\!\!\!\mathop{=}\limits^{{\rm (d)}}\frac{1}{N}I(S_{3}^{N};S_{1}^{N}|S_{2}^{N})+\frac{1}{N} (H(Y_{3}^{N}|X_{2}^{N},Y_{2}^{N})-}
\\[.1cm] {{\ }H(Y_{3}^{N}|X_{1}^{N},X_{2}^{N},Y_{2}^{N}))+\varepsilon_{1}+\varepsilon}
\\[.1cm] {\!\!\!=\frac{1}{N} I(S_{1}^{N} ;S_{3}^{N} \left|S_{2}^{N} )\right. +\frac{1}{N} I(X_{1}^{N} ;Y_{3}^{N} \left|X_{2}^{N} ,Y_{2}^{N} )\right. +\varepsilon _{1} +\varepsilon }
\\ [.1cm]{\!\!\!\le I(S_{1} ;S_{3} \left|S_{2} )\right. +I(X_{1} ;Y_{3} \left|X_{2} ,Y_{2} )\right. +\varepsilon _{1} +\varepsilon }
\end{array}$
}
\normalsize{

\noindent where (a) results from the security condition, (b) from Fano's inequality, (c) from the fact that given $(S_{1} ,S_{2})$, $S_{3}$ is independent of $(X_{1},X_{2},Y_{2})$ and (d) from the fact that given $(X_{1},X_{2})$, $Y_{3}$ is independent of $(S_{1},S_{2},S_{3})$.
}

The sum rate upper bound can be deduced from Fano's equality.
\vspace{-.1cm}
\noindent
\section*{Appendix III}

\noindent
\subsection*{Proof of the converse in Theorem 2}

\noindent To derive the outer bound in Theorem 2, for an arbitrary small $\varepsilon>0$, we use Fano's inequality and security conditions as:
\small{
\[\begin{array}{l}
{H(K_{1},K_{2}|Y_{3}^{N},S_{3}^{N})\le N(\frac{H(\varepsilon)}{N}+\varepsilon(\log |{\rm{\mathcal K}}_{{\rm 1}}||{\rm {\mathcal K}}_{{\rm 2}} |-1))\triangleq N\varepsilon_{1},}
\\[.2cm]{I(K_{1} ;S_{2}^{N} ,X_{2}^{N} ,Y_{2}^{N} )<N\varepsilon, {\, \, }I(K_{2} ;S_{1}^{N} ,X_{1}^{N} ,Y_{1}^{N} )<N\varepsilon,}
\end{array}\]
}
\normalsize{

\noindent where $\varepsilon _{1}\to0$ if $\varepsilon\to0$.}
We prove the outer bound of $R_{1}$. The outer bound of $R_{2}$ can be deduced by symmetry.
\vspace{-.2cm}
\small{
\[\begin{array}{l} {\!\!\!R_{1} =\frac{1}{N} H(K_{1} )\mathop{\le }\limits^{{\rm (a)}} \frac{1}{N} H(K_{1} \left|S_{2}^{N} ,X_{2}^{N} ,Y_{2}^{N} )\right. +\varepsilon}
 \\{\!\!\!\mathop{\le }\limits^{{\rm (b)}} \frac{1}{N} H(K_{1} \left|S_{2}^{N} ,X_{2}^{N} ,Y_{3}^{N} )\right. +\varepsilon } \\{\!\!\!\mathop{\le }\limits^{{\rm (c)}} \frac{1}{N} H(K_{1} \left|S_{2}^{N} ,X_{2}^{N} ,Y_{3}^{N} )\right. -\frac{1}{N} H(K_{1} \left|S_{3}^{N} ,Y_{3}^{N} )\right. +\varepsilon _{1} +\varepsilon }
 \\[.1cm]{\!\!\!\le \frac{1}{N}H(K_{1}\!|S_{2}^{N}\!,X_{2}^{N}\!,Y_{3}^{N}\!)\!-\frac{1}{N}H(K_{1}\!|S_{2}^{N}\!,X_{2}^{N},S_{3}^{N},Y_{3}^{N})\!+\varepsilon _{1}\!+\!\varepsilon}
 \\{\!\!\!\mathop{=}\limits^{{\rm (d)}}\!\frac{1}{N}H(K_{1}\!|S_{2}^{\!N}\!,X_{1}^{\!N}\!,\!X_{2}^{\!N}\!,\!Y_{3}^{N\!}\!)\!-\frac{1}{N}H(K_{1}\!|S_{2}^{\!N}\!,\!X_{1}^{\!N}\!,\!X_{2}^{\!N}\!,\!S_{3}^{\!N} \!,\!Y_{3}^{\!N}\!)\!+\!\varepsilon_{1}\!+\!\varepsilon}
 \\{\!\!\!\mathop{=}\limits^{{\rm (e)}}\frac{1}{N}H(K_{1}\!|S_{2}^{\!N}\!,\!X_{1}^{\!N}\!,\!X_{2}^{\!N}\!)\!-\!\frac{1}{N}H(K_{1}\!|S_{2}^{\!N}\!,\!X_{1}^{\!N}\!,X_{2}^{\!N}
 \!,\!S_{3}^{\!N}\!)\!+\!\varepsilon_{1}\!+\!\varepsilon}
 \\[.1cm]{\!\!\!\le \frac{1}{N} H(K_{1} \left|S_{2}^{N} ,X_{1}^{N} )\right. -\frac{1}{N} H(K_{1} \left|S_{2}^{N} ,X_{1}^{N} ,S_{3}^{N} )\right. +\varepsilon _{1} +\varepsilon }
\\[.1cm]{\!\!\!=\!\frac{1}{N}I(\!K_{1};\!S_{3}^{\!N}\!|S_{2}^{\!N}\!,\!X_{1}^{N}\!)\!+\!\varepsilon_{1}\!+\!\varepsilon\!\le\! \frac{1}{N}I(K_{1} ,X_{1}^{N};S_{3}^{N}|S_{2}^{N})\!+\!\varepsilon_{1}\!+\!\varepsilon}
 \\[.1cm]{\!\!\!=\frac{1}{N} \sum _{i=1}^{N}I(K_{1} ,X_{1}^{N} ;S_{3,i} \left|S_{3,i+1}^{N} ,S_{2}^{N} )\right. +\varepsilon _{1} +\varepsilon  {\rm \; }}
 \\{\!\!\!\mathop{=}\limits^{{\rm (f)}} \frac{1}{N} \sum _{i=1}^{N}I(K_{1} ,X_{1}^{N} ;S_{3,i} \left|S_{3,i+1}^{N} ,S_{2,1}^{i} )\right. +\varepsilon _{1} +\varepsilon {\rm \; } }
 \\[.1cm]{\!\!\!{\rm =}\frac{1}{N}\!\sum_{i=1}^{N}{\rm\! [\!} H(S_{3,i}\!|S_{3,i+1}^{\!N}\!,S_{2,1}^{i}\!)\!-\!H(S_{3,i} \!|S_{3,i+1}^{\!N} \!,S_{2,1}^{i}\!,K_{1}\!,X_{1}^{\!N}\!)]\!+\!\varepsilon_{1}\!+\!\varepsilon}
 \end{array}\]}
 \vspace{-.2cm}
\[\begin{array}{l}{\!\!\!\le \frac{1}{N} \sum_{i=1}^{N}{\rm\![}H(S_{3,i}\!|S_{2,i}\!)\!-\!H(\!S_{3,i}\!|S_{3,i+1}^{N}\!,\!S_{2,1}^{i} \!,\!S_{1,1}^{i-1},K_{1},X_{1}^{N}\!)]\!+\!\varepsilon_{1}\!+\!\varepsilon\!}
 \\{\!\!\!\mathop{=}\limits^{{\rm (g)}}\!\sum_{i=1}^{N}{\rm [}H(S_{3,i}\!|S_{2,i}\!)\!-\!H\!(\!S_{3,i}\!|S_{3,i+1}^{N} \!,S_{2,i}^{}\!,\!S_{1,1}^{i-1}\!,K_{1}\!,X_{1}^{N}\!)]\!+\!\varepsilon_{1}\!+\!\varepsilon\!}
 \\{\!\!\!\mathop{=}\limits^{{\rm (h)}} \frac{1}{N} \sum _{i=1}^{N}I(U_{1,i} ;S_{3,i} \left|S_{2,i} )+\varepsilon _{1} +\varepsilon \right.  \mathop{=}\limits^{{\rm (i)}} I(U_{1,Q} ;S_{3,Q} \left|S_{2,Q} )\right. +\varepsilon '{\rm \; }} \end{array}\]}
\normalsize{
\vspace{-.2cm}

\noindent where (a) results from the security condition, (b) from the Markov chain $(\!X_{1}\!,\!X_{2}\!)\!-\!\!Y_{2}\!-\!\!Y_{3}$, (c) from Fano's inequality, (d) from the fact that $H(\!X_{1}\!|X_{2}\!,\!Y_{3}\!)=0$, (e) from the fact that given $(X_{1}\!,\!X_{2})$, $Y_{3}$ is independent of $(\!S_{1}\!,\!S_{2}\!,\!S_{3}\!)$,(f) from the fact that given $S_{3,i+1}^{\!N}$,$(K_{1}\!,X_{1}^{\!N} )$ is independent of $S_{2,i+1}^{\!N}$ which is the direct consequence of the Markov chain $S_{1}\!-\!S_{3}\!-\!S_{2}$ , (g) from the Markov chain $S_{3,i}\!-\!(\!S_{3,i+1}^{N},\!S_{2,i},\!S_{1,1}^{i-1},\!K_{1},\!X_{1}^{N}\!)\!-\!S_{2,1}^{i-1}$ which arises because $(K_{1},X_{1}^{N})$ is a function of $S_{1}^{N}$, (h) from the definition of the random variable $U_{1,i}$ as:
\vspace{-.1cm}
\begin{equation}
 U_{1,i}=(S_{3,i+1}^{N},S_{1,1}^{i-1},K_{1},X_{1}^{N})\no
 \end{equation}

\vspace{-.05cm}
\noindent and (i) from the definition of the random variable $Q$ which is uniformly distributed on ${\rm\{\!}1,\!2,...,\!N{\rm\}\!}$ and setting $\varepsilon'\!\!=\!\!\varepsilon_{1}\!\!+\!\!\varepsilon$. With the same approaches as above, we have $R_{2}\!\!\le\!\! I(U_{2,Q};\!S_{3,Q}\!|S_{1,Q})\!+\!\varepsilon'$
where $U_{2,i}$ is defined as $U_{2,i}\!=\!(S_{3,1}^{i-1},\!S_{2,i+1}^{N},\!K_{2},\!X_{2}^{N})$. It can be seen that the defined random variables satisfy the distribution of theorem 2.
}

To prove the rate constraints of Theorem 2, we have:

\vspace{.2cm}
\small{
\noindent $\begin{array}{l} {NI(X_{1}^{} ;Y_{3}^{} \left|X_{2}^{} )\right. \ge I(X_{1}^{N} ;Y_{3}^{N} \left|X_{2}^{N} )\right. \ge I(S_{1}^{N} ;Y_{3}^{N} \left|X_{2}^{N} )\right. }
\\ {\mathop{\ge }\limits^{{\rm (a)}} I(S_{1}^{N} ;Y_{3}^{N} \left|X_{2}^{N} )\right. +H(K_{1} \left|S_{3}^{N} ,Y_{3}^{N} ,X_{2}^{N} )\right. -N\varepsilon _{1} }
\\ [.1cm]{\ge I\!(\!S_{1}^{\!N}\!;Y_{3}^{\!N}\!|X_{2}^{\!N}\!)\!+\!H(K_{1}\!|S_{3}^{\!N}\!,Y_{3}^{N}\!,X_{2}^{N}\!)\!-\!H(K_{1} \!|S_{1}^{\!N}\!,Y_{3}^{N}\!,X_{2}^{\!N}\!)\!-\!N\varepsilon_{1}\!}
\\ [.1cm]{=I(S_{1}^{N} ;Y_{3}^{N} ,K_{1} \left|X_{2}^{N} )\right. -I(S_{3}^{N} ;K_{1} \left|Y_{3}^{N} ,X_{2}^{N} )\right. -N\varepsilon _{1} }
\\ [.1cm]{\ge I(S_{1}^{N} ;Y_{3}^{N} ,K_{1} \left|X_{2}^{N} )\right. -I(S_{3}^{N} ;Y_{3}^{N} ,K_{1} \left|X_{2}^{N} )\right. -N\varepsilon _{1} }
\\ {\mathop{=}\limits^{{\rm (b)}} I(S_{1}^{N} ;Y_{3}^{N} ,K_{1} ,X_{1}^{N} \left|X_{2}^{N} )\right. -I(S_{3}^{N} ;Y_{3}^{N} ,K_{1} ,X_{1}^{N} \left|X_{2}^{N} )\right. -N\varepsilon _{1} }
\\ {\mathop{=}\limits^{{\rm (c)}} I(S_{1}^{N} ;K_{1} ,X_{1}^{N} \left|X_{2}^{N} )\right. -I(S_{3}^{N} ;K_{1} ,X_{1}^{N} \left|X_{2}^{N} )\right. -N\varepsilon _{1} }
\\ {\mathop{=}\limits^{{\rm (d)}} H(K_{1} ,X_{1}^{N} \left|S_{3}^{N} )\right. -I(K_{1} ,X_{1}^{N} \left|S_{1}^{N} )\right. -N\varepsilon _{1} }
\\[.1cm]{=\!I(\!S_{1}^{N} ;K_{1}\!,X_{1}^{N}\!)\!-\!I(S_{3}^{N} ;K_{1} ,X_{1}^{N} )-N\varepsilon _{1} }
\\[.1cm]{\!=\!I(\!K_{1}\!,X_{1}^{\!N}\!;S_{1}^{\!N}\!|S_{3}^{N}\!)\!-\!N\varepsilon_{1}\!=\!\sum _{i=1}^{N}I\!(\!K_{1}\!,X_{1}^{\!N};\!S_{1,i}\!|S_{3}^{\!N}\!,S_{1,1}^{i-1}\!)\!-N\varepsilon_{1}}
\\[.1cm]{=\sum _{i=1}^{N}I(K_{1} ,X_{1}^{N} ;S_{1,i} \left|S_{3,i}^{N} ,S_{1,1}^{i-1} )\right. -N\varepsilon _{1}  }
\\[.1cm] {=\sum _{i=1}^{N}[H(S_{1,i} \left|S_{3,i} )\right. -H(S_{1,i} \left|S_{3,i}^{N} ,S_{1,1}^{i-1} ,K_{1} ,X_{1}^{N} )]\right. -N\varepsilon _{1}  }
\\[.1cm] {=\sum _{i=1}^{N}I(K_{1} ,X_{1}^{N} ,S_{3,i+1}^{N} ,S_{1,1}^{i-1} ;S_{1,i} \left|S_{3,i} )\right. -N\varepsilon _{1}}  \\{=I(U_{1,Q} ;S_{1,Q} \left|S_{3,Q} )\right. -N\varepsilon _{1}.}
\end{array}$
}
\vspace{.1cm}

\normalsize{
In the above equations, (a) results from Fano's inequality, (b) from the fact that $H(\!X_{1}|X_{2},\!Y_{3}\!)\!=\!0$, (c) from the fact that given $(X_{1},\!X_{2})$, $Y_{3}$ is independent of $(S_{1},S_{2},S_{3})$ and (d) from the Markov chain \small{$(S_{1}^{N},\!K_{1},\!X_{1}^{N}\!)\!-\!S_{3}^{N}\!-\!(S_{2}^{N},\!K_{2},\!X_{2}^{N})$}.

\normalsize{
\indent With the same approaches, we have $N\!I(X_{2};\!Y_{3}|X_{1})\!\!\ge \!\!\!I(U_{2,Q};S_{2,Q}|S_{3,Q})\!-\!N\varepsilon_{1}$}}.
\vspace{.1cm}

\noindent For the sum rate constraint, we have:
\vspace{.1cm}

\small{
\noindent $\begin{array}{l} {\!\!\!\sum_{i=1}^{N}I(U_{1,i},U_{2,i};S_{1,i},S_{2,i}|S_{3,i})}
\\[.1cm]{\!\!\!=\!\sum_{i=1}^{N}I(K_{1},X_{1}^{N},S_{3,i+1}^{N},\!S_{1,1}^{i-1},K_{2},X_{2}^{N},S_{3,1}^{i-1},S_{2,i+1}^{N};S_{1,i} ,S_{2,i}|S_{3,i})}
\\[.1cm]{\!\!\!=\!\sum_{i=1}^{N}[H(S_{1,i},S_{2,i}|S_{3,i})-}
\\[.1cm]{{\ }H(S_{1,i},S_{2,i}|K_{1},X_{1}^{N},S_{1,1}^{i-1},K_{2},X_{2}^{N},S_{2,i+1}^{N},S_{3}^{N})]}
\\[.1cm] {\!\!\!=\!\sum_{i=1}^{N}[H(S_{1,i},S_{2,i}\!|S_{3,i})-H(S_{1,i}|K_{1},X_{1}^{N},S_{1,1}^{i-1},S_{3}^{N})-}
\\[.1cm]{{\ }H(S_{2,i}\!|K_{2},X_{2}^{N},S_{2,i+1}^{N},S_{3}^{N})]}
\\[.1cm]{\!\!\!=H(S_{1}^{N},S_{2}^{N}|S_{3}^{N})-H(S_{1}^{N}|K_{1},X_{1}^{N},S_{3}^{N})-H(S_{2}^{N} |K_{2},X_{2}^{N},S_{3}^{N})}
\\[.1cm] {\!\!\!\le H(S_{1}^{N},S_{2}^{N}|S_{3}^{N})-H(S_{1}^{N}|K_{1},X_{1}^{N},X_{2}^{N},Y_{3}^{N},S_{3}^{N})-}
\\[.1cm]{{\ \ }H(S_{2}^{N}|K_{2},X_{1}^{N},X_{2}^{N},Y_{3}^{N},S_{3}^{N})}
\end{array}$

\noindent $\begin{array}{l} 
{\!\!\!=\!H(S_{1}^{N},S_{2}^{N}|S_{3}^{N})-H(S_{1}^{N}|K_{1},X_{2}^{N},Y_{3}^{N},S_{3}^{N})-}
\\[.1cm]{{\ }H(S_{2}^{N} |K_{2},X_{1}^{N},Y_{3}^{N},S_{3}^{N})}
\\ {\!\!\!\mathop{=}\limits^{{\rm (a)}} H(S_{1}^{N} ,S_{2}^{N} \left|S_{3}^{N} )-H(S_{1}^{N} \left|K_{1} ,Y_{3}^{N} ,S_{3}^{N} )-H(S_{2}^{N} \left|K_{2} ,Y_{3}^{N} ,S_{3}^{N} )\right. \right. \right. }
\\[.1cm] {\!\!\!\le I(S_{1}^{N} ,S_{2}^{N} ;K_{1} ,K_{2} ,Y_{3}^{N} |S_{3}^{N} )}
\\[.1cm]{\!\!\!=H(K_{1} ,K_{2} ,Y_{3}^{N} \left|S_{3}^{N} )-H(K_{1} ,K_{2} ,Y_{3}^{N} \left|S_{3}^{N} ,S_{1}^{N} ,S_{2}^{N} )\right. \right. }
\\ {\!\!\!\mathop{\le }\limits^{{\rm (b)}} H(Y_{3}^{N} \left|S_{3}^{N} )-H(Y_{3}^{N} \left|S_{3}^{N} ,S_{1}^{N} ,S_{2}^{N} )\right. \right. +N\varepsilon _{1} }
\\ [.1cm]{\!\!\!\le H(Y_{3}^{N} )-H(Y_{3}^{N} \left|X_{1}^{N} ,X_{2}^{N} ,S_{3}^{N} ,S_{1}^{N} ,S_{2}^{N} )\right. +N\varepsilon _{1} }
\\ [.1cm]{\!\!\!=H(Y_{3}^{N} )-H(Y_{3}^{N} \left|X_{1}^{N} ,X_{2}^{N} )\right. +N\varepsilon _{1} =I(X_{1}^{N} ,X_{2}^{N} ;Y_{3}^{N} )+N\varepsilon _{1} }
\\[.1cm]{\!\!\!\le NI(X_{1}^{} ,X_{2}^{} ;Y_{3}^{} )+N\varepsilon _{1} }
\end{array}$
}

\normalsize{
In the above equations, (a) results from the Markov chain $(S_{1}^{N}\!,K_{1}\!,X_{1}^{N})\!-S_{3}^{N}\!-(S_{2}^{N},\!K_{2},\!X_{2}^{N})$ and (b) from Fano's inequality.
}
%
%
%
%
%

}
%
%

\begin{thebibliography}{1}
%
%
%
%
\bibitem{Ahlswede} R. Ahlswede and I. Csisz´ar, \lq\lq Common randomness in information theory and cryptography, part I: Secret sharing,\rq\rq ~\emph{IEEE Trans. Inf. Theory}, vol. 39, no. 4, pp. 1121--1132, Jul. 1993.
\bibitem{Maurer} U. M. Maurer, \lq\lq Secret key agreement by public discussion from common information,\rq\rq ~\emph{IEEE Trans. Inf. Theory}, vol. 39, no. 3, pp. 733--742, May 1993.
\bibitem{helper} I. Csiszar, P. Narayan, \lq\lq Common randomness and secret key generation with a helper,\rq\rq ~\emph{IEEE Trans. Inf. Theory}, vol. 46, no. 2, pp.344-366, Mar 2000.
\bibitem{joint-source-channel} A. Khisti, S. Diggavi, G. Wornell, \lq\lq Secret key generation using correlated sources and noisy channels,\rq\rq ~\emph{IEEE Int.Symp. Inf. Theory}, Toronto, Canada, pp. 1005-1009, Jul. 2008.
\bibitem{salimi-new-source-model} S. Salimi, M. Salmasizadeh, M. R. Aref, \lq\lq Secret Key Sharing in a New Source Model: Rate Regions,\rq\rq ~\emph{IET Communications}, Vol. 5, Issue 4, pp. 443-455, March 2011.
\bibitem{mac-poor} Y. Liang and V. Poor, \lq\lq Multiple access channels with confidential messages,\rq\rq ~\emph{IEEE Trans. Inf. Theory}, vol. 54, no. 3, pp. 976-1002, Mar. 2008.
\bibitem{Wyner-Ziv} M. Gastpar, \lq\lq The Wyner-Ziv problem with multiple sources,\rq\rq ~\emph{IEEE Trans. Inf. Theory}, vol. 50, no. 11, pp. 2762 - 2768, Nov. 2004.

\end{thebibliography}
\end{document}